\documentclass[twocolumn,prb]{revtex4}
\usepackage{epsfig}
\usepackage{mathrsfs}
\usepackage{amssymb}
\usepackage{color}
\usepackage{latexsym}
\usepackage{subfigure}
\newcommand{\bcen}{\begin{center}}
\newcommand{\ecen}{\end{center}}
\newcommand{\btab}{\begin{tabular}}
\newcommand{\etab}{\end{tabular}}
\newcommand{\bdes}{\begin{description}}
\newcommand{\edes}{\end{description}}

\newcommand{\beq}{\begin{equation}}
\newcommand{\eeq}{\end{equation}}
\newcommand{\bea}{\begin{eqnarray}}
\newcommand{\eea}{\end{eqnarray}}

\newcommand{\etal}{et~al.\ }
\newcommand{\half}{\frac{1}{2}}
\newcommand{\bary}{\begin{array}}
\newcommand{\eary}{\end{array}}
\newcommand{\benum}{\begin{enumerate}}
\newcommand{\eenum}{\end{enumerate}}
\newcommand{\bitem}{\begin{itemize}}
\newcommand{\eitem}{\end{itemize}}
\newcommand{\eqn}[1] {eqn.~(\ref{#1})}

\newcommand{\fig}[1]{FIG.~\ref{#1}}
\newcommand{\Fig}[1]{FIG.~\ref{#1}}
\newcommand{\addn}[1]{{\color[rgb]{0,0,0}{#1}}}

\newcommand{\mycaption}[1]{\caption[#1]{#1}}
\newcommand{\mean}[1]{\langle #1 \rangle}

\makeatletter
\def\@dotsep{4.5}
\makeatother

\begin{document}
\title{Edge State Magnetism of Single Layer Graphene Nanostructures}
\author{Somnath~Bhowmick$^1$}\email[]{bsomnath@mrc.iisc.ernet.in}
\author{Vijay B Shenoy$^{2,1}$}\email[]{shenoy@physics.iisc.ernet.in}
\affiliation{$^1$Materials Research Center, Indian Institute of Science, Bangalore 560 012, India\\
$^2$Centre for Condensed Matter Theory, Department of Physics, Indian Institute of Science, Bangalore 560 012, India}
\date{\today}
%ABSTRACT-------------------------------------------------------------
\begin{abstract}

We study edge state magnetism in graphene nanostructures using a
mean field theory of the Hubbard model. We investigate how the
magnetism of the zigzag edges of graphene is affected by the presence of other
types of terminating edges and defects. By a detailed study of both
regular shapes, such as polygonal nanodots and nanoribbons, and
irregular shapes, we conclude that the magnetism in zigzag edges is
very robust. Our calculations show that the zigzag edges that are longer
than three to four repeat units are always magnetic, irrespective of
other edges, regular or irregular. We, therefore, clearly demonstrate
that the edge irregularities and defects of the bounding edges of graphene
nanostructures does not destroy the edge state magnetism. 
\end{abstract}
%--------------------------------------------------------------------------
\maketitle
%INTRODUCTION----------------------------------------------------------------

\section{Introduction}
\label{intro}
Graphene is a single layer of carbon atoms forming a densely packed
honeycomb lattice.  Novoselov \etal~\cite{novoselov2004,novoselov2005}
invented the top-down technique of isolating single layered graphene
samples of a few microns in size by micromechanical cleavage of
graphite. This advance has lead to a flurry of activity, both
theoretical and experimental, towards understanding the physics of
graphene.\cite{geim2007,katsnelson2007,neto2007}

 A simple tight binding model\cite{wallace1947} for electron
hopping on the honeycomb lattice produces two bands which touch each other at two points in the
Brillouin zone. The chemical potential of the undoped graphene lies
exactly at the energy where the two bands touch, implying that graphene is
a zero gap system with two distinct ``Fermi points''.  What makes it
even more interesting is the fact that the spectrum near these Fermi
points resembles the Dirac spectrum of massless
Fermions,\cite{semenoff1984,halden1988,novoselov_nat2005} and the
density of states depends linearly on the energy. Graphene shows
several interesting magnetotransport properties, such as quantum Hall
effect at room temperature,\cite{novoselov2007} unconventional
quantum Hall effect,\cite{novoselov2006} strongly suppressed weak
localization magentoresistance\cite{morozov2006} and quantum
electrodynamics phenomena such as Klein paradox~\cite{katsnelson2006}
etc. Due to it's unusual electronic properties, graphene is a
strong contender for future electronic applications.  Examples are,
graphene based field effect transistors (FET),\cite{katsnelson2006}
spin valve devices,\cite{hill2006} gas sensors,\cite{schedin2007}
integrated ballistic carrier devices based on nanopatterned epitaxial
graphene~\cite{berger2004} etc.

\begin{figure*}
\subfigure[]{\epsfxsize=6.0truecm \epsfbox{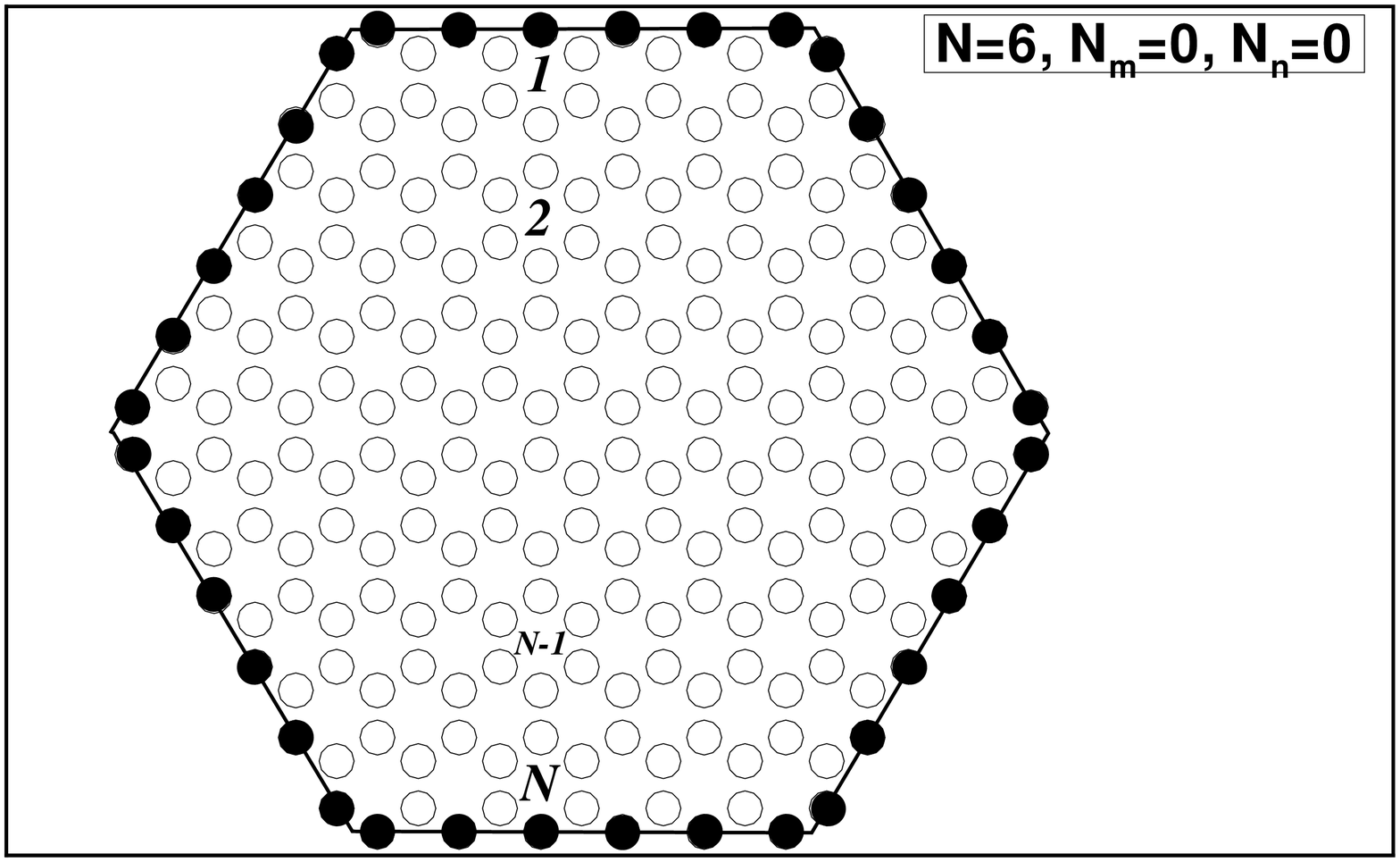}}
\subfigure[]{\epsfxsize=6.0truecm \epsfbox{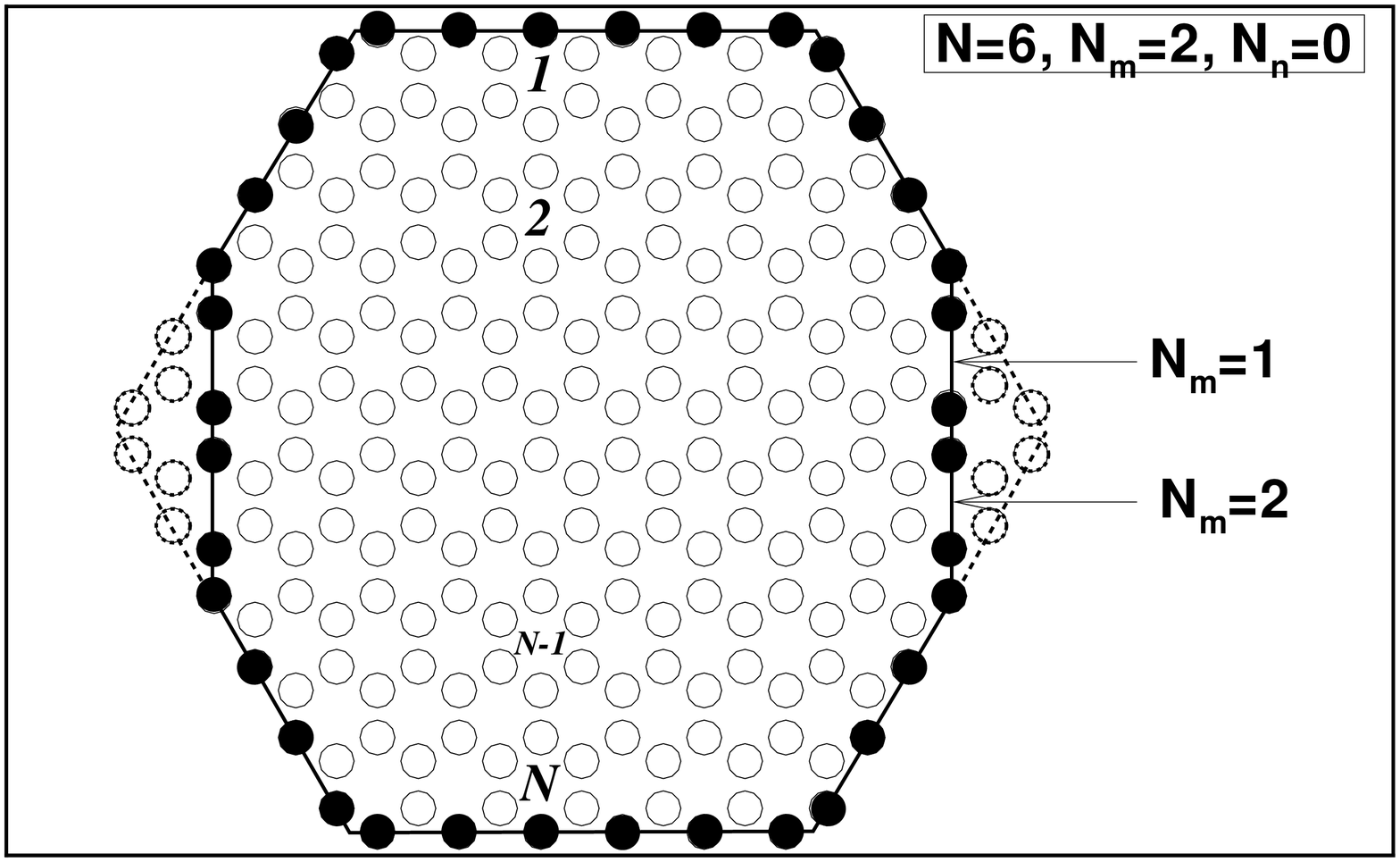}}
\subfigure[]{\epsfxsize=6.0truecm \epsfbox{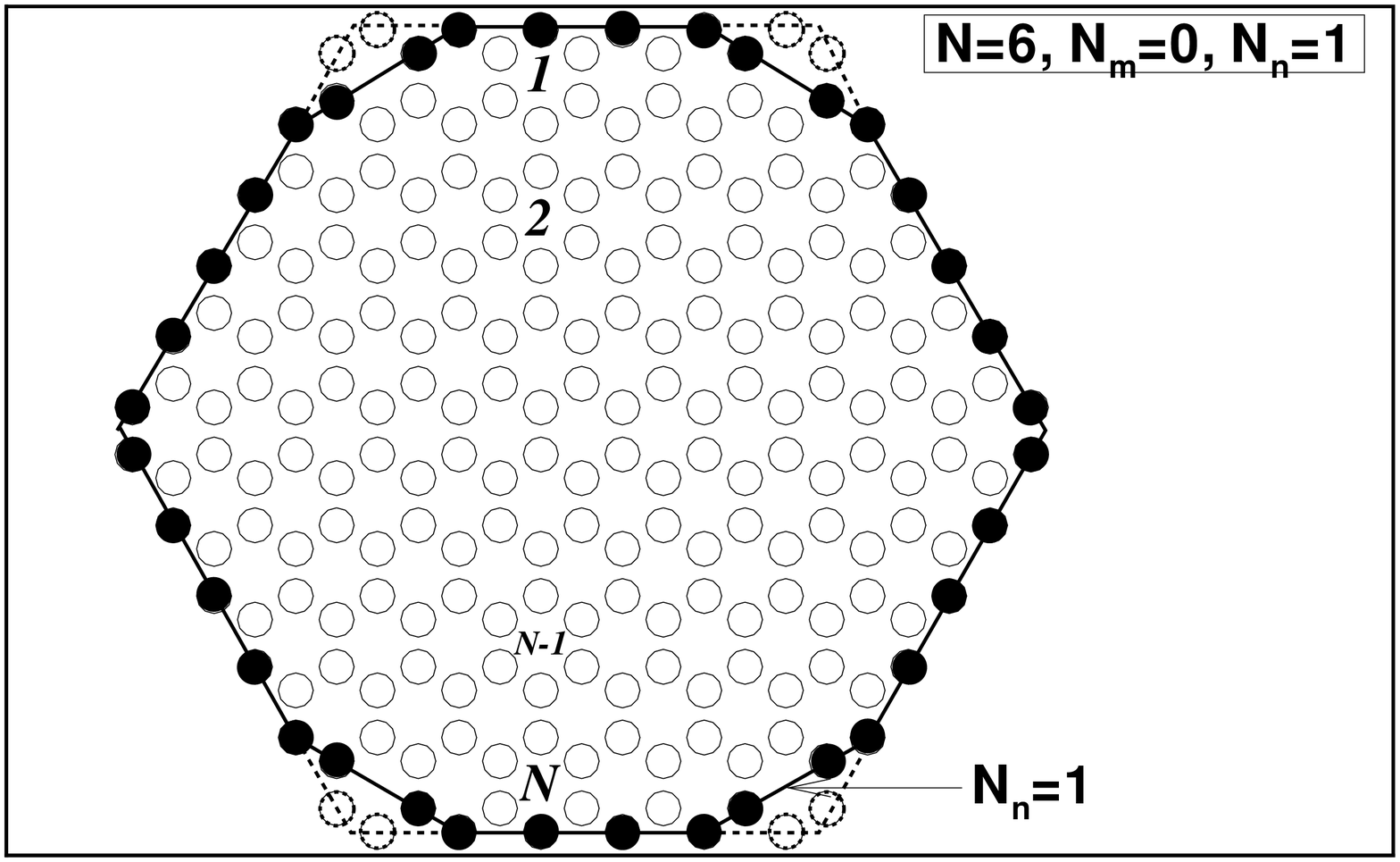}}
\subfigure[]{\epsfxsize=6.0truecm \epsfbox{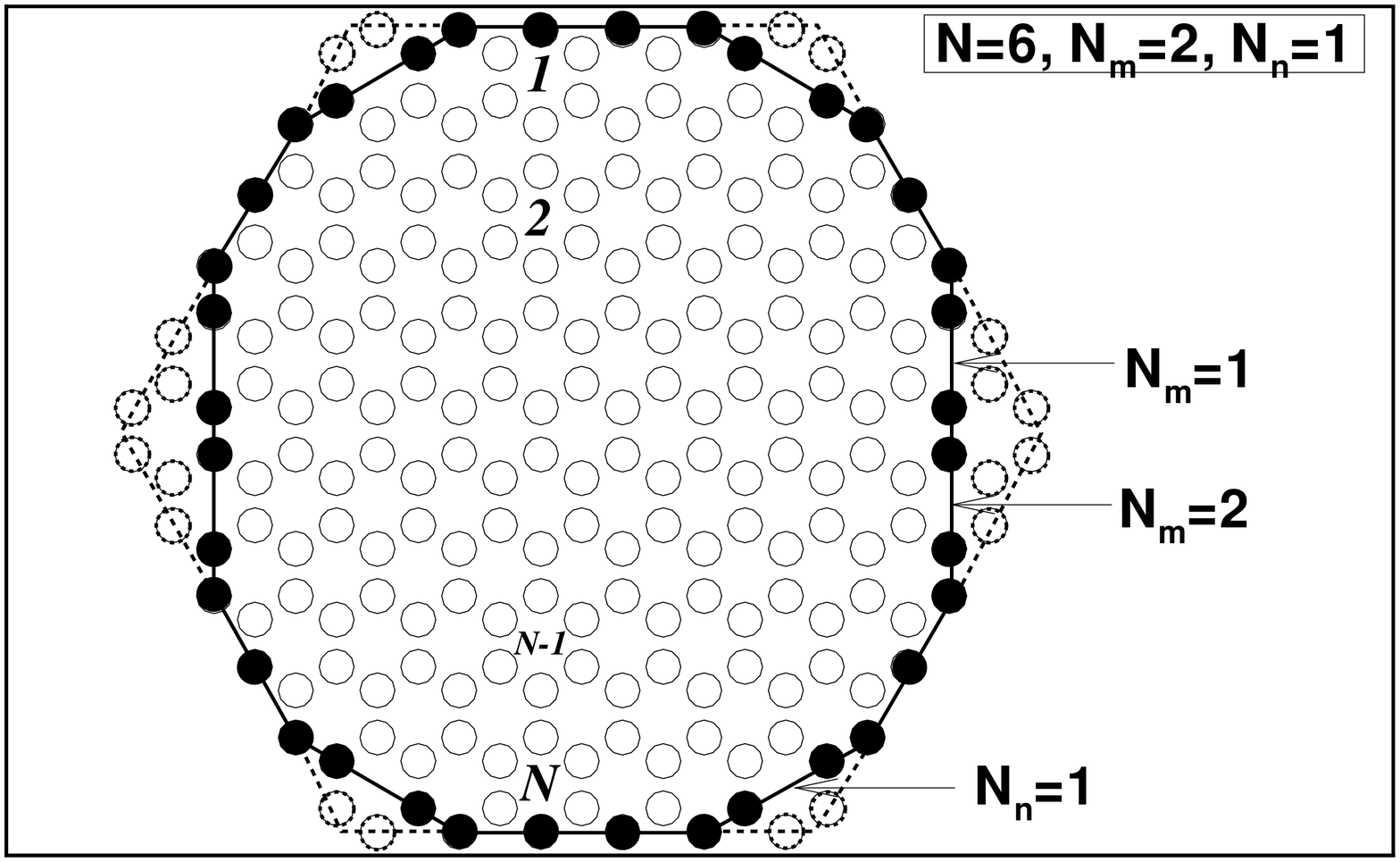}}
\mycaption{  (a) Hexagon: all the edges are of zigzag type. The size of the nanodot
is denoted by $N$. We cut this polygon in different ways in order to
produce various other polygons having both armchair and zigzag
edges. Dark circles represent the edge sites in all the figures, and
the dashed circles represent atoms that are removed. (b) Eight edged
polygon: created by vertically cutting the hexagon resulting in two
additional armchair edges per vertical side (quantified by $N_m$). The dotted portion
in (b), (c) and (d) is the remaining part of the original full
hexagon.  (c) Ten edged polygon: obtained by cutting the hexagon
along the slanted edge (no vertical cut) resulting in one armchair edge per slanted side
(quantified by $N_n$) in the process. (d)
Twelve edged polygon: created by simultaneously cutting along the
vertical as well as the slanted edges, adding total four armchair
edges of each type along the vertical and slanted directions respectively.
}
\label{nanodots}
\end{figure*}

Interesting nanostructures can also be made from graphene. For
example, a nanoribbon\cite{nakada1996} is obtained by
reducing the dimension of a graphene sheet along one direction to the
nanometric size, and a nanodot by reducing both the dimensions to the
nanometric sizes.  Graphene nanostructures can be terminated by many
different types of edges\cite{nakada1996}, for example, by ``zigzag''
or ``armchair'' edges (see \Fig{nanodots}). These edges can have a
profound influence on the electronic structure, and can give rise to
interesting new phenomena.  For example, zigzag edges have localized
electronic states with nearly flat dispersion, giving rise to the finite DOS
at the chemical potential, as has been
reported~\cite{katsuyoshi1993,nakada1996} from the theoretical
calculations. These have also been experimentally observed by scanning
tunneling microscopy (STM) and
spectroscopy~\cite{kobayashi2005,niimi2006}. Indeed, the presence of the
zigzag edges gives rise to unique physical and chemical properties as
reported by Son \etal~\cite{young2006} and Jiang
\etal~\cite{jiang_jcp2007} based on the first principle density
functional calculations.

%fig======================================================================
%============================================================================

Focus of the present paper is on the edge state magnetism 
of graphene nanostructures. Based on first principle
density functional calculations, it has been reported that the zigzag
nanoribbons have an antiferromagnetic (AFM) ground state, with one edge
spin up and the other spin down~\cite{okada2001}, whereas the armchair
edged nanoribbons do not show any such magnetic property. \addn{The
investigation has further been extended to graphene nanodots of
rectangular shape~\cite{jiang2007} and regular hexagonal, as well as triangular nanodots~\cite{rossier2007}
by first principle and mean field Hubbard model calculations; the results of these two approaches are  found to be in agreement.\cite{rossier2007}}

All the reports so far deal with the nanodots
terminated by consecutive zigzag edges. However, the widely used
top-down technique~\cite{novoselov2004} does not allow a careful
control of the synthesis at the atomic level to produce graphene
nanodots of regular shape, terminated, for example, entirely by zigzag
edges. Furthermore, in spite of recent advances~\cite{wu2007},
bottom-up techniques which are likely to give better control on the
resulting structure are not yet widely practiced.
The simplest defects to the ``ideal'' zigzag edged nanostructures are armchair terminations.\cite{oleg2008}
Despite the fact that there are several studies of defect induced magnetism in graphene,
\cite{oleg2008,oleg_cm2008,oleg2007} to the best of our knowledge, edge
state magnetism of {\it realistic}
graphene nanostructures, in particular those randomly terminated by the zigzag or
armchair edges is yet to be examined in detail. In this work we address this issue and understand how the
magnetism of zigzag edges ``copes'' with the presence of other edges
and random terminations.  Using an unrestricted Hartree-Fock mean-field
theory of the Hubbard model on the graphene lattice, we study
nanostructures with various different terminations with the aim of
uncovering how the edge magnetism is affected.  Three cases are
investigated. First, we study nanodots by starting from a perfect
hexagonal shape and systematically introducing other types of
edges. Second, we investigate graphene nanoribbons terminated by the
zigzag edges and systematically introduce ``surface edge defects'' and
study the resulting magnetism. Third, we study ``random''
nanostructures by introducing completely irregular edges, while
preserving some zigzag segments at the edges.
\addn{ The Hartree-Fock approach allows us to study large nanostructures with as many as 1000 carbon atoms; such a study within the first principles framework would require much bigger computational resources. However, we do not take into account effects like bond length variations and
reconstructions at the edges, as well as effects of functional groups (e.g. oxidized edges) in
the present calculation.}

The remainder of the paper is organized as follows. In the next
section we brief the mean field Hubbard model used in our study. This is
followed by a section (Sec.~\ref{result}) where we present and discuss
our results. In the concluding section (Sec.~\ref{conclusion}), we
discuss the significance of our results particularly in the context of
experimental observation of edge magnetism in graphene.

%fig======================================================================
\begin{figure*}
\subfigure[]{\epsfxsize=6.0truecm \epsfbox{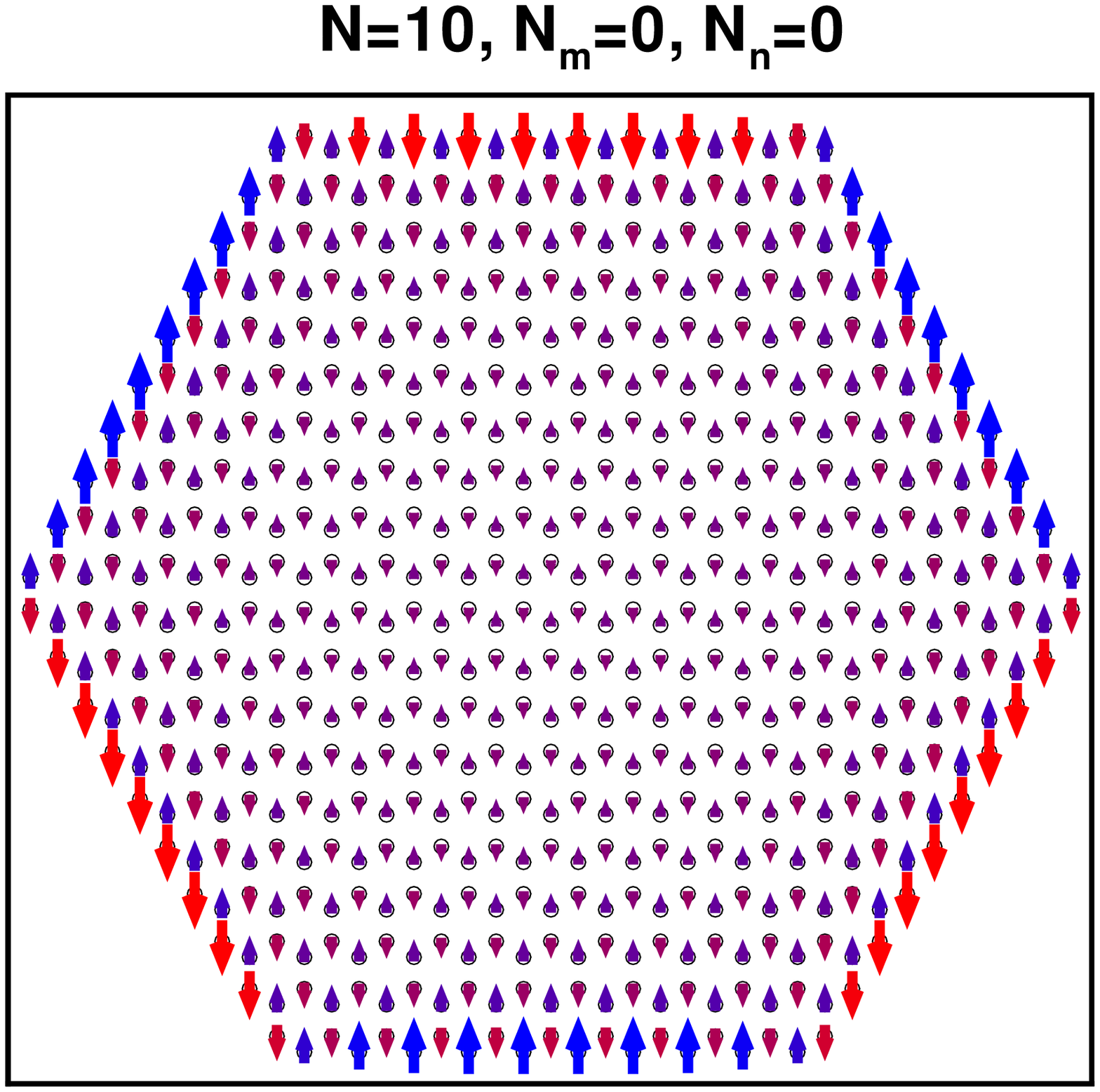}}
\subfigure[]{\epsfxsize=6.0truecm \epsfbox{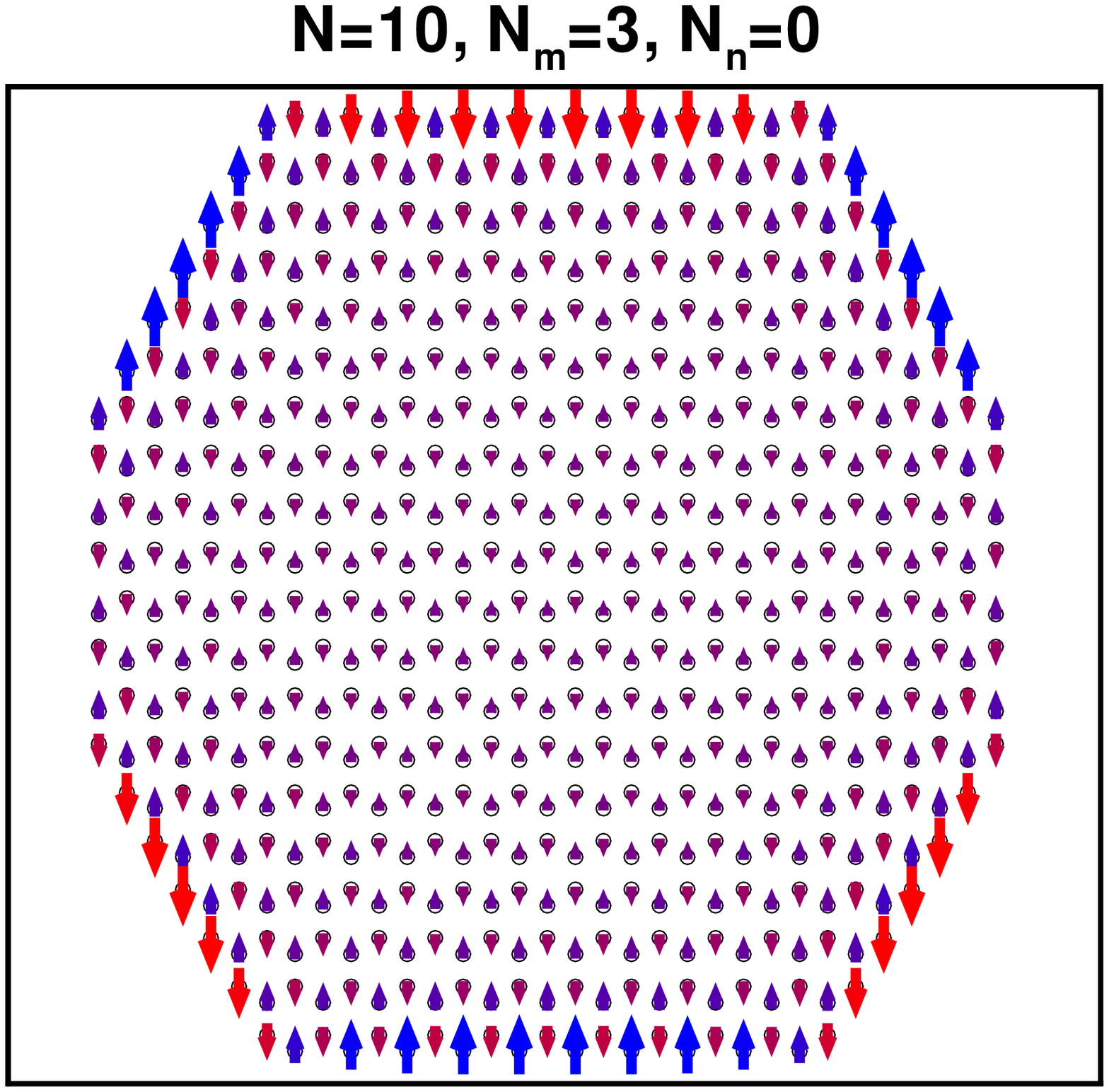}}
\subfigure[]{\epsfxsize=6.0truecm \epsfbox{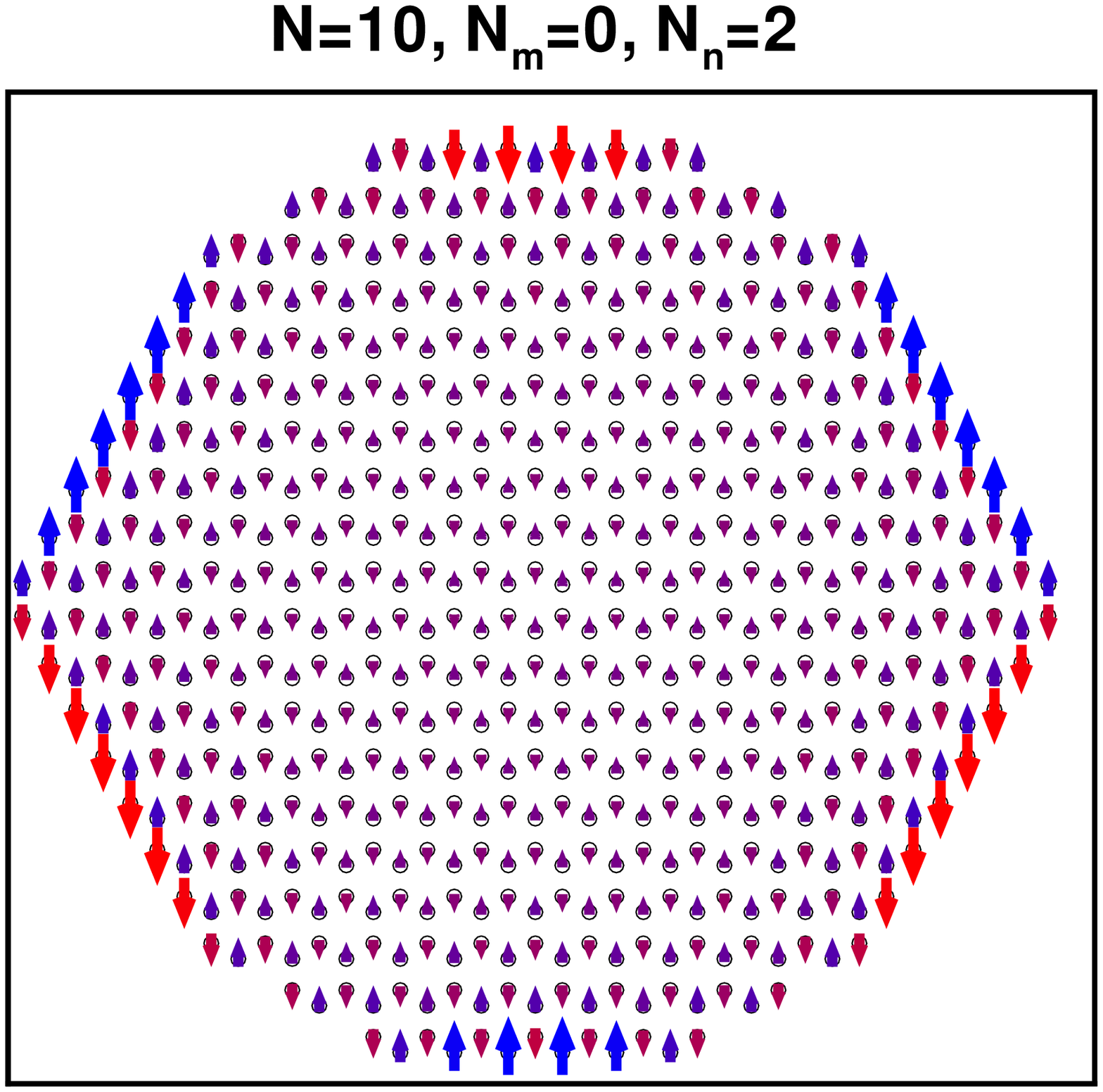}}
\subfigure[]{\epsfxsize=6.0truecm \epsfbox{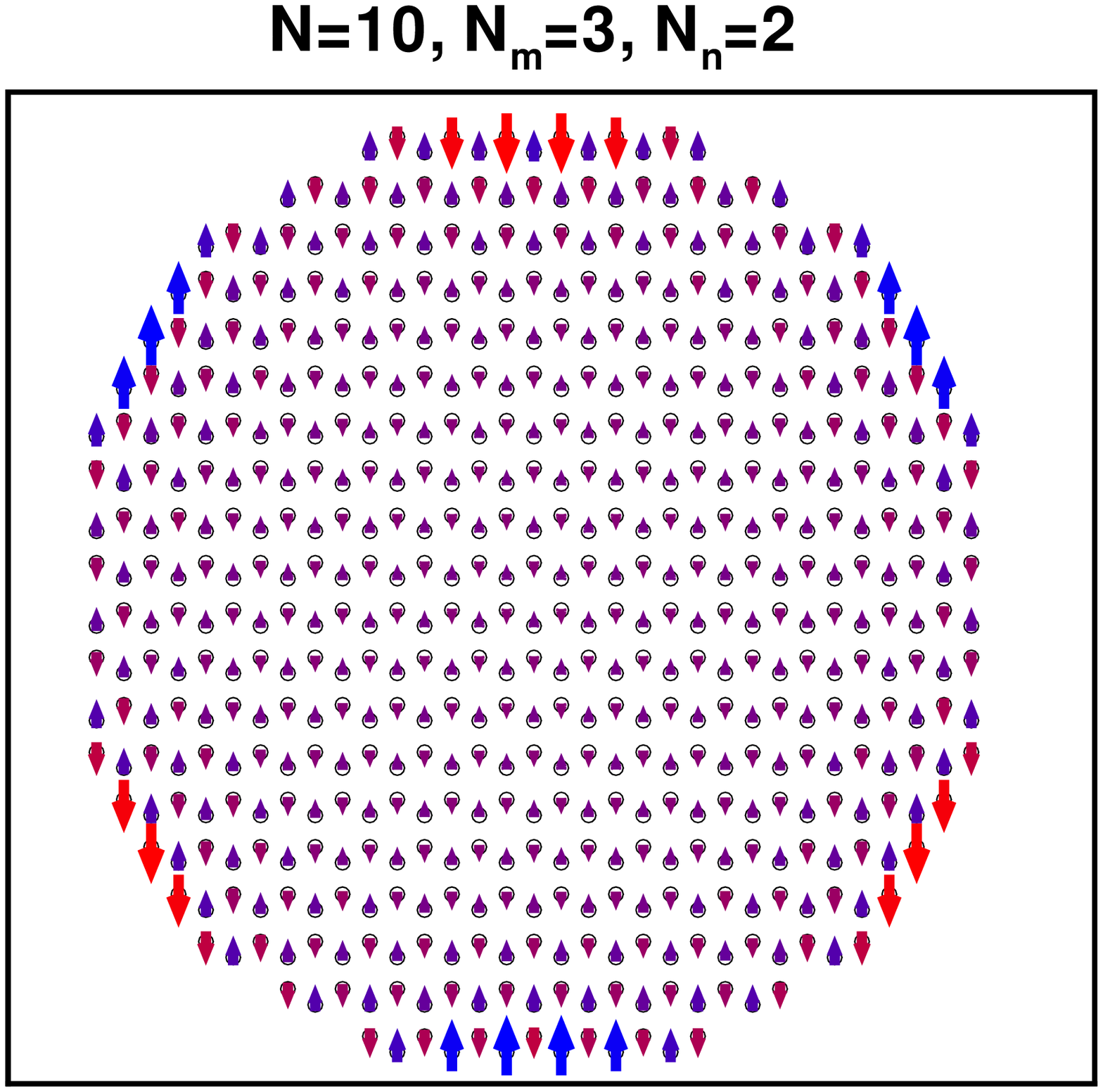}}
\mycaption{   (Online Color) Expectation value of $\mean{S^z}$ in four different graphene nanodots. The 
size of the arrows is proportional to the magnitude of magnetic moment. Note that there exist large
magnetic moments at the zigzag edges compared to the sites lying inside or armchair edges.}
\label{nanodotmoment}
\end{figure*}
%===========================================================================

\section{Model Details}
\label{model}

We study graphene within a tight binding model including the on-site Hubbard repulsion, i.~e., within the framework of the Hubbard model\cite{Baskaran2002,Black-Schaffer2007,rossier2007}, which has been used earlier to study surface magnetism.\cite{potthoff1995,potthoff1997} The Hamiltonian is
\bea
H = -t\sum_{\langle ij\rangle\sigma}(c^{\dagger}_{i\sigma}c_{j \sigma}+h.c) + U \sum_{i} n_{i\uparrow} n_{i\downarrow}
\eea
where $c_{i\sigma}(c^\dagger_{i\sigma})$ annihilates(creates) an
electron of spin $\sigma$ at site $i$, $n_{i \sigma} =
c^\dagger_{i\sigma} c_{i\sigma}$ is the number operator at site $i$,
$t$ is the hopping amplitude, $U$ is the on-site Coulomb
repulsion. The Hamiltonian has lattice translational and SU(2) spin rotation symmetries. 
Under the mean field approximation, we write the Hubbard
Hamiltonian as
\begin{equation}
H=-t\sum_{\langle ij\rangle\sigma}(c^{\dagger}_{i\sigma}c_{j\sigma}+h.c)+U\sum_i(n_{i\uparrow}\langle n_{i\downarrow}
\rangle
+n_{i\downarrow}\langle n_{i\uparrow}\rangle-\langle n_{i\downarrow}\rangle\langle n_{i\uparrow}\rangle)
\end{equation}
where $\mean{n_{i \sigma}}$ are the site occupations which are
determined self consistently. We perform a completely unrestricted
calculation keeping all site occupations $\mean{n_{i \sigma}}$ as
unknowns. We focus on the undoped graphene, which had exactly one electron
per site, i.~e., we work with the half filled Hubbard model. From the
self consistent ground state calculated numerically, we obtain the
expectation value of $z$-component of the spin operator
${S_{i}^z} = \half c^{\dagger}_{i\alpha}\sigma^z_{\alpha \alpha'} c_{i \alpha'}$ to study the resulting
magnetic structure ($\sigma^z$ is a Pauli matrix).

We first discuss known results\cite{furukawa2001} of the magnetic
phases of the half filled Hubbard model on an infinite honeycomb
lattice. The honeycomb lattice does not break any symmetry of the
Hamiltonian until the onsite Coulomb repulsion attains a critical
value\cite{furukawa2001} of $U_{c}/t\sim 2.23$. Above this value of the onsite Coulomb
repulsion, the system breaks SU(2) spin rotation and lattice
translational symmetries and develops an antiferromagnetic order (due
to its bipartite nature). This is unlike the 2D square lattice (with the nearest neighbour hopping),
where  even an infinitesimally small $U$ will give rise to the antiferromagnetic
spin density wave(SDW)~\cite{furukawa2001} at half filling. This result of the honeycomb lattice
may be understood starting from the magnetic susceptibility $\chi(q)$ obtained from the linear
response theory,\cite{longbook}
\bea
\label{lindhard}
\chi(q)=(-1)\frac{1}{N}\sum_k\frac{f(\varepsilon_k-\mu)-f(\varepsilon_{k+q}-\mu)}{\varepsilon_k-\varepsilon_{k+q}}
\eea
where $f(\varepsilon)$ is the Fermi function and $q$ is the wave
vector of perturbing field, $\varepsilon_{k}$ is the electron dispersion, $\mu$ is the chemical potential. Now the half filled Hubbard model on a  bipartite lattice with nearest
neighbour hopping has particle-hole symmetry. Using  \eqn{lindhard},
the bare magnetic susceptibility (for the antiferromagnetic response)
can be obtained as\cite{longbook,Mahan2000}
\beq
\chi_0 \sim \int d\varepsilon\frac{\rho(\varepsilon)}{\varepsilon}
\label{suscep}
\eeq
where $\rho(\varepsilon)$ is the density of states(DOS). In case of a square lattice,
$\rho(\varepsilon)\sim log(\varepsilon)$ in the vicinity of chemical potential at zero doping and $\chi_0$
has a logarithmic divergence. As a consequence, the  generalized Stoner criterion  for anti ferromagnetic
(AF) susceptibility,\cite{longbook}
\bea
U\ge\frac{1}{\chi_0}
\label{sc}
\eea
is satisfied for even an infinitesimally small $U$ and $U_c=0$ for square
lattice. However, in case of a honeycomb lattice, DOS depends linearly on the energy measured
with respect to the chemical potential~\cite{peres2004},
\beq
\rho(\varepsilon)\sim \varepsilon
\label{dos}
\eeq
Hence the divergence of $\chi_0$ is absent (consult \eqn{suscep}),
i.e., $\chi_0$ is a finite quantity. Antiferromagnetism, therefore,
appears only for Coulomb repulsion greater than a critical value
$U_c=1/\chi_0= 2.23 t$.  Experimentally large graphene sheets do not
show any magnetic ordering and hence the value of $U$ in graphene
should be less than $U_c$ (this result, of course, does not include
fluctuational corrections). We shall use a value of $U = 2 t$ in our
calculations, which at the mean field level will not produce any
magnetic order in an infinite graphene sheet.

As noted in the introductory section, zigzag terminated nanoribbons (quasi 1D)
possess special edge states which give rise to the flat bands and
finite density of states at the Fermi
level.\cite{nakada1996,katsuyoshi1993,castro2008} This leads to ferromagnetic spin
orientations along a particular zigzag edge at infinitesimally small $U_c$ (from \eqn{suscep}).
On the other hand, two opposite zigzag edges, terminating the nanoribbon at two opposite sides,
are found to be aligned antiferromagnetically. 
The spin density (magnitude of the magnetic moment at a
site) dies very quickly on moving into the ``bulk'', i.e., the centre
of the nanoribbon -- this is indeed edge state
magnetism. Armchair edges do not support any magnetic structure for
values of $U$ below $U_c$, in contrast to zigzag edges.

In the next section we shall investigate various graphene
nanostructures terminated by zigzag and other edges with the aim of
investigating how the presence of other edges affects the edge state
magnetism.

%fig======================================================================
\begin{figure}[h]
\centerline{\epsfxsize=6.0truecm \epsfbox{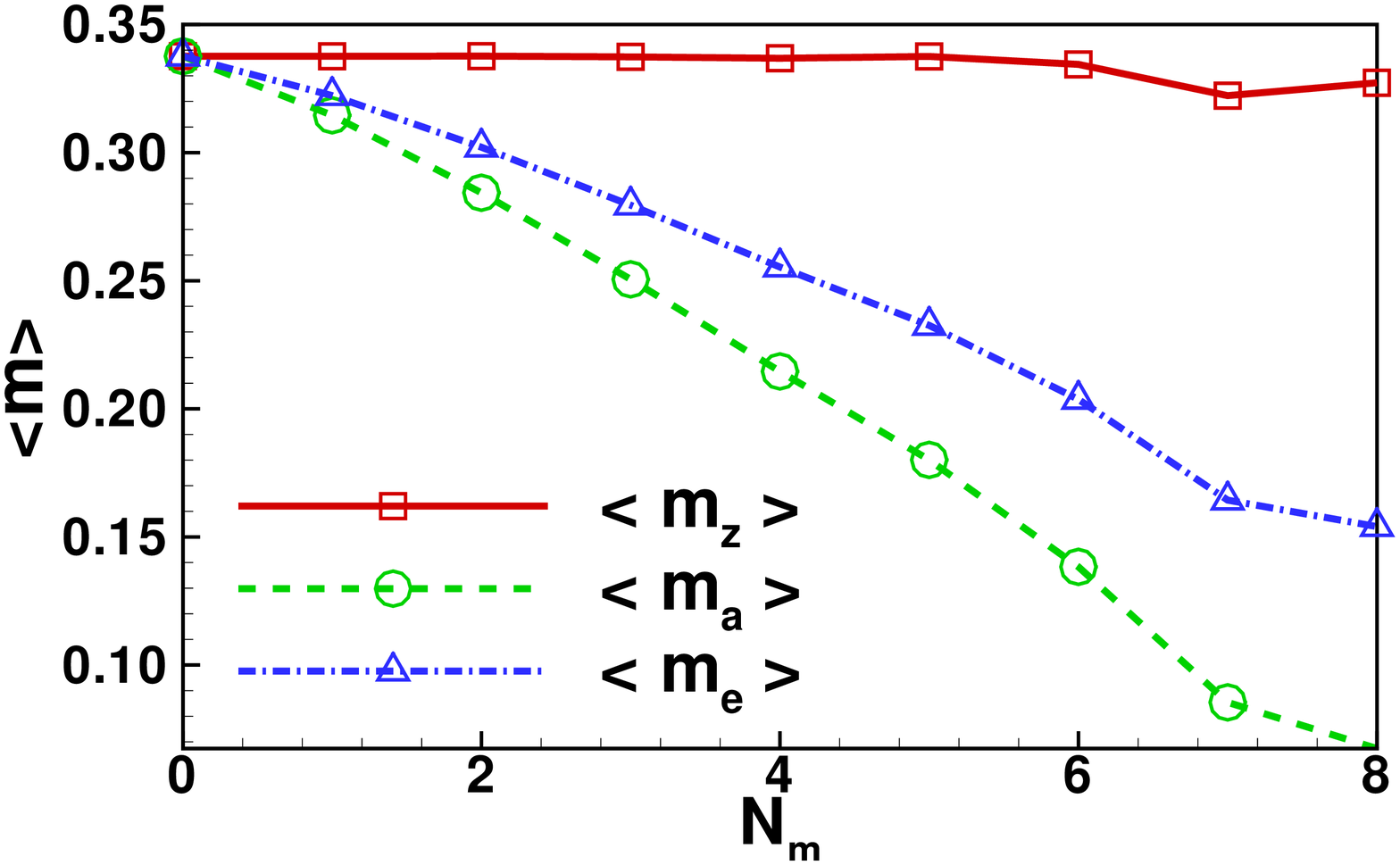}}
\mycaption{(Online Color)    Average magnetic moment per edge site for the eight edged polygons. $N_m$
is the number of armchair edges added along the vertical side (\fig{nanodots}(b)).  $\langle m_z
\rangle$ is the average magnitude of the magnetic moment per site
along the purely zigzag (horizontal sides in \fig{nanodots}(b) and
\fig{nanodotmoment}(b)) edges. $\langle m_a \rangle$ is the average
magnitude of the magnetic moment per site along the non-horizontal
edges. $\langle m_e \rangle$ is the magnitude of the magnetic moment per site averaged
over all the edge sites (zigzags as well as armchairs). As we increase $N_m$, we add
armchairs at the non-horizontal sides and hence
$\langle m_a \rangle$, as well as $\langle m_e \rangle$,
decreases. Moreover, in the process of increasing $N_m$, we do not
disturb the purely zigzag horizontal edges and $\langle m_z \rangle$
remains almost unaffected.
}
\label{meight}
\end{figure}
%=================================================================================

\section{Graphene Nanostructures: Edge State Magnetism}
\label{result}

We study edge state magnetism in three types of nanostructures. Regular nanodots with zigzag and
armchair edges, nanoribbons with ``defected'' zigzag edges, and random nanostructures with some
zigzag segments. \addn{Before discussing each nanostructure in detail, we comment on the
nature of the expected ground state.  Total spin $S_z$ of the ground
state of regular nanodots is found\cite{rossier2007} to be zero and
this result is consistent with the predictions of Lieb's
theorem.~\cite{lieb1989} This arises from the fact that the earlier
calculations with regular nanodots will have equal number of atoms of
the ``A-sublattice'' and the ``B-sublattice''.  On the other hand for
irregular nanostructures, $N_A$ need not be equal to $N_B$, for
example see \fig{rand2}(c) and (d) and total $S_z$ is expected (and found) to be nonzero in such
cases.}

\subsection{Nanodots}

Starting from a perfect hexagonal
nanodot,\cite{rossier2007} which is entirely
enclosed by zigzag edges (\fig{nanodots}(a)), we can make various
different polygons, that are terminated by both zigzags and armchair
edges (see \fig{nanodots}(b), (c),(d)). In this process we retain
zigzag edges on the horizontal sides of the resulting
polygon. Nomenclature of size etc.~is explained
in~\fig{nanodots}. The sizes of the horizontal zigzag segments can by controlled
by varying the lengths of the armchair segments $N_m$ and $N_n$.

%fig======================================================================
\begin{figure}
\centerline{\epsfxsize=6.0truecm \epsfbox{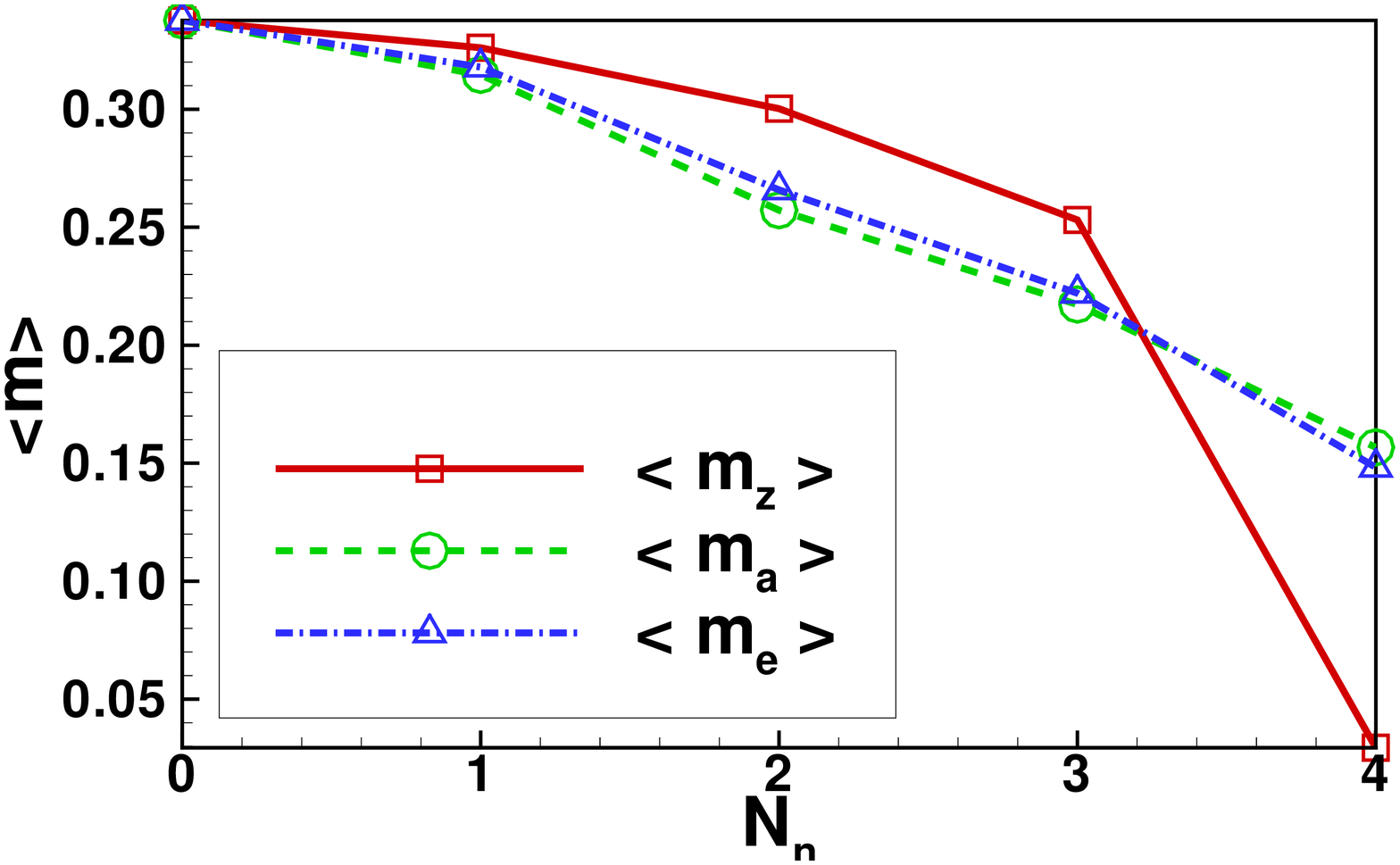}}
\caption{(Color online)
Average magnetic moment per edge site for the ten edged polygons. $N_n$ is the number of armchair
edges added along the
slanted sides of the hexagonal polygon (\fig{nanodots}(c)). Consult \fig{meight} for the definition
of $\langle m_z \rangle$, $\langle m_a \rangle$ and $\langle m_e\rangle$. Overall edge magnetization
diminishes with the increase of number of armchairs. In particular, as we increase $N_n$
{\it the horizontal zigzag segments are reduced in length} and the most notable change occurs for
$N_n$ greater than three, where the magnetization of the horizontal edges ($\langle m_z \rangle$)
essentially vanishes. Note that for $N_n=3$ there are four repeat units in the horizontal
zigzag segment and for $N_n=4$ the zigzag segments are two repeat units long.}
\label{mten}
\end{figure}
%======================================================================

The magnetic structure obtained from the mean field analysis of the
Hubbard model is shown in \fig{nanodotmoment}. We observe that the
surface sites, in particular those along the zigzag edges but not along the
armchair edges, have large expectation values of the $\mean{S_z}$
operator. Moreover, the magnetic moment decreases sharply as we move
towards the centre of the nanodot and at the ``bulk'' sites, it is in
general at least one to two order of magnitude smaller in comparison to that
at the zigzag edge sites. In
all cases we find that the moment is along the ``up direction'' on
sites of one sub-lattice and ``down direction'' on the sites belonging
to the other sub-lattice. We shall henceforth focus on the magnitude
of the magnetic moment, keeping in mind this observation.

%fig======================================================================
\begin{figure}
\subfigure[]{\epsfxsize=6.0truecm \epsfbox{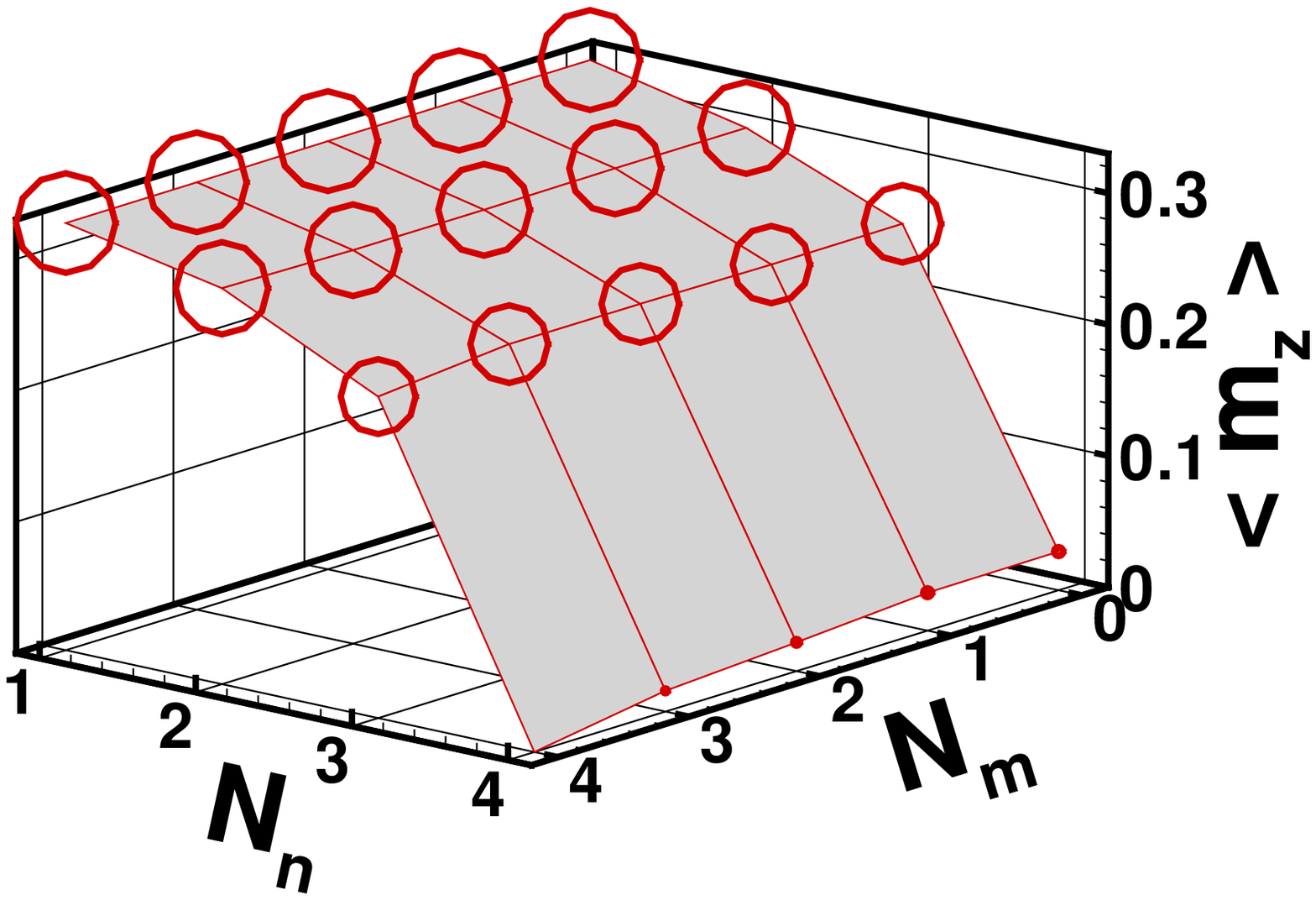}}
\subfigure[]{\epsfxsize=6.0truecm \epsfbox{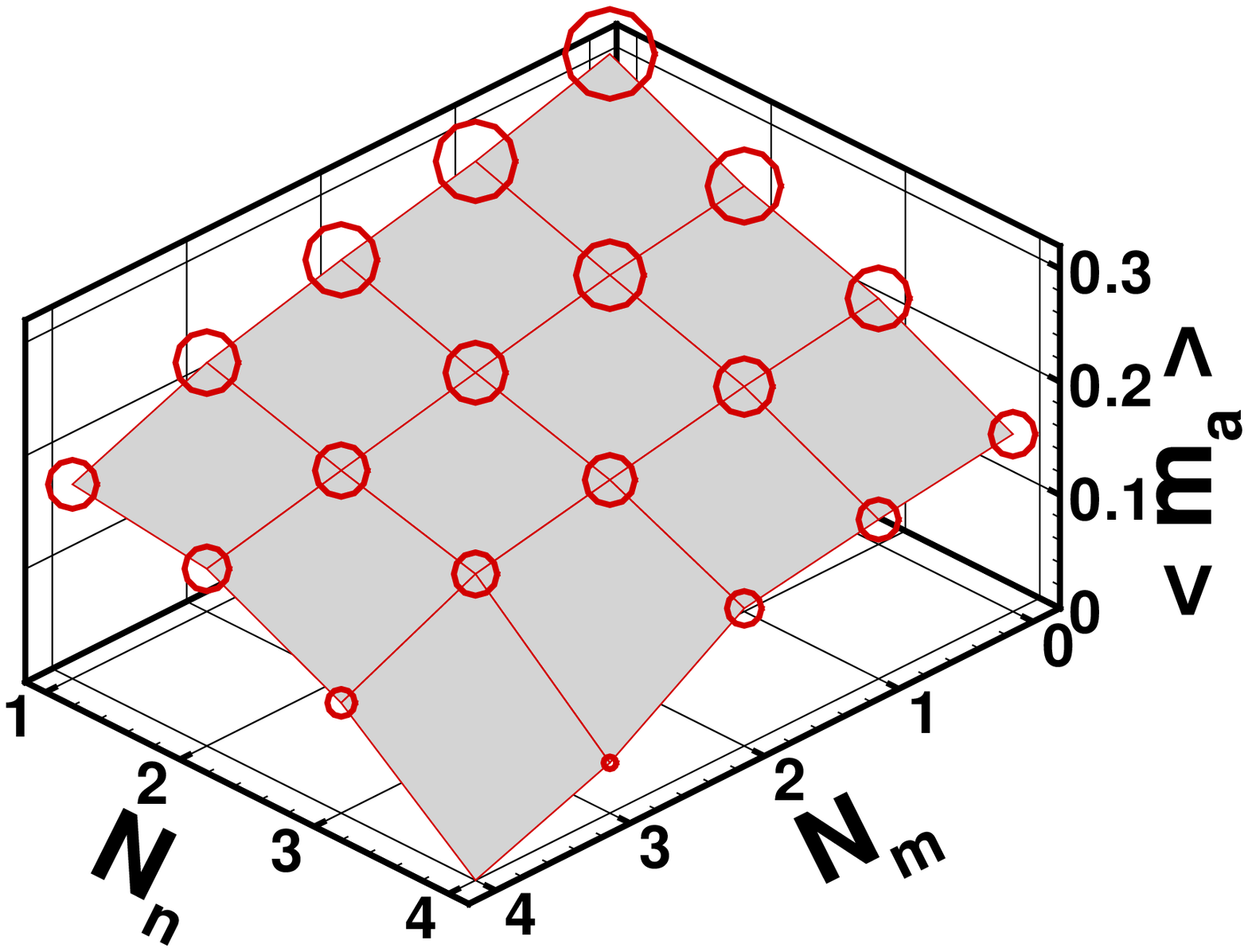}}
\caption{(Color online) Average magnetic moment per edge site for the twelve edged nanodots
(\fig{nanodots}(d)). See \fig{meight}
for the definition of $\langle m_z \rangle$ and $\langle m_a \rangle$. The size of
the symbols is proportional to the magnitude of the average magnetic moment. (a) The key finding is that
the magnetic moment along the horizontal zigzag edges is affected by only their length and not
by the presence of other edges. Increase of $N_m$ does not affect the length of the horizontal
zigzag edges, while increase of $N_n$ decreases the length of the horizontal edges. (b)
$\langle m_a \rangle$, average magnitude of magnetic moment along the non-horizontal sides decreases
as we increase $N_m$ and $N_n$, thereby adding more armchairs to the edges.}
\label{mtwelve}
\end{figure}
%===============================================================================

We now present a quantitative analysis of the how the magnetism at the
horizontal zigzag edges is affected by the introduction of other
segments. \Fig{meight},~\fig{mten} and \fig{mtwelve} show the plots of the
magnitude of the magnetic moments per site for the three types of regular
polygons constructed from a hexagonal nanodot discussed in
\Fig{nanodots}. The average magnitude of the magnetic moment
per site along the zigzag segments is denoted by $\mean{m_z}$, while
$\mean{m_a}$ denotes the average of the magnitude of the magnetic
moment obtained on non-horizontal edges (the non-horizontal edges
contain a mixture of both zigzag and armchair edges depending of $N_m$
and $N_n$ ). The average magnitude of the magnetic moment over all the
edge sites is denoted by $\mean{m_e}$.  \Fig{meight} shows the
variation of various average magnetic moments for the eight edged
nanostructure of shown in \Fig{nanodots}(b). We find that the
magnetization of the horizontal zigzag edges is unaffected by the
presence of additional non-horizontal edges.  The average magnitude of
the magnetization $\mean{m_e}$ along all edges falls due to the
presence of additional armchair segments. In \Fig{mten}, we show the
results of average magnetization for the ten sided polygons. In this
case the addition of non-horizontal armchair edges results (i.~e.,
increase of $N_n$) in a reduction in the length of the horizontal
zigzag segments. In this case we see that the magnetization of the
zigzag segments falls (and so does the average magnetization) and for
$N_n > 3$, it essentially vanishes. The case of $N_n = 3$
corresponds to having horizontal zigzag segments whose length is four
repeat units. We turn now to the twelve sided polygons shown in
\Fig{nanodots}(d) where we introduce two types nonhorizontal armchair
segments denoted by $N_m$ and $N_n$. As in the case of the eight sided
polygon the length of the horizontal zigzag segment is unaffected by
increase of $N_m$ while the length of the horizontal zigzag segments
is reduced up on increase of $N_n$. We see that the magnetization of
the horizontal edges is unaffected by increase of $N_m$, but strongly
affected by increase of $N_n$ which reduces the length of the
horizontal zigzag edges. The upshot of our calculation is that the
zigzag edges retain a significant amount of magnetization even in the
presence of other edges as long as the length of the zigzag segments
is greater than three to four repeat units. The number of repeat units of the zigzag edges that preserve
the magnetization depends on the value of $U$. For example, our calculations show that for
$U=1.2t$, the number of repeat units must be five to six for retaining the magnetization.
%fig======================================================================
\begin{figure*}
\subfigure[]{\epsfxsize=6.0truecm \epsfbox{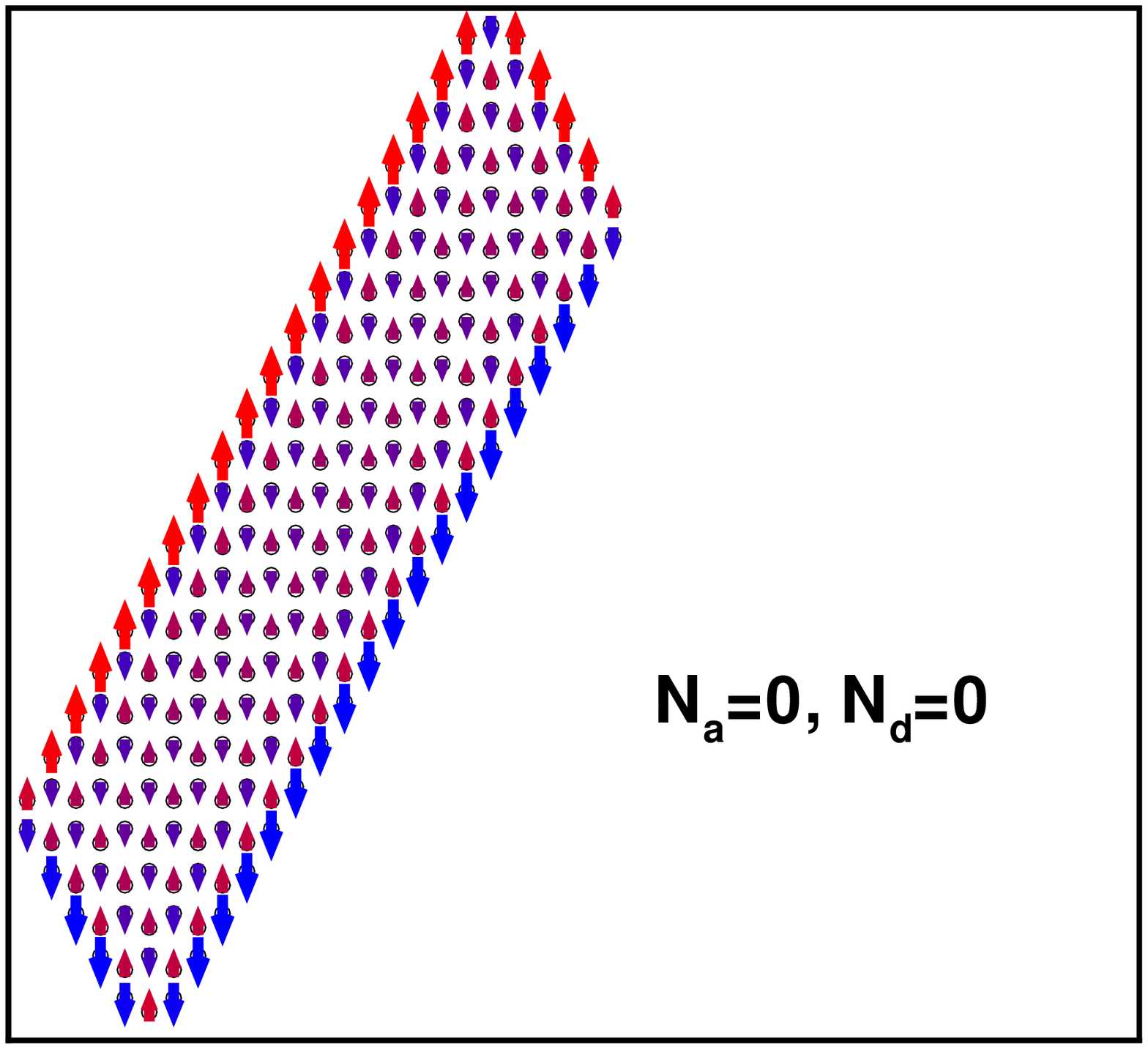}}
\subfigure[]{\epsfxsize=6.0truecm \epsfbox{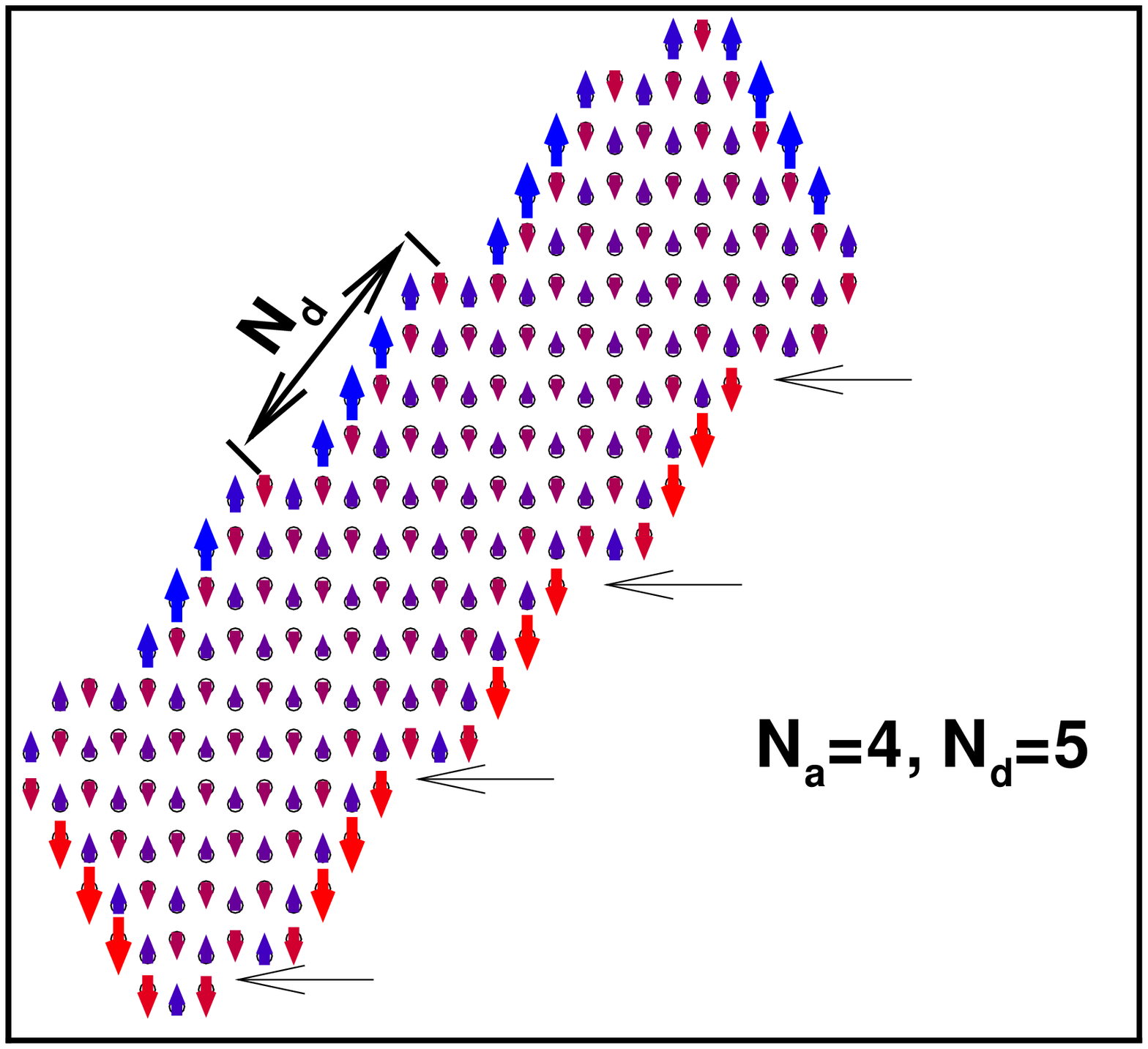}}
\subfigure[]{\epsfxsize=6.0truecm \epsfbox{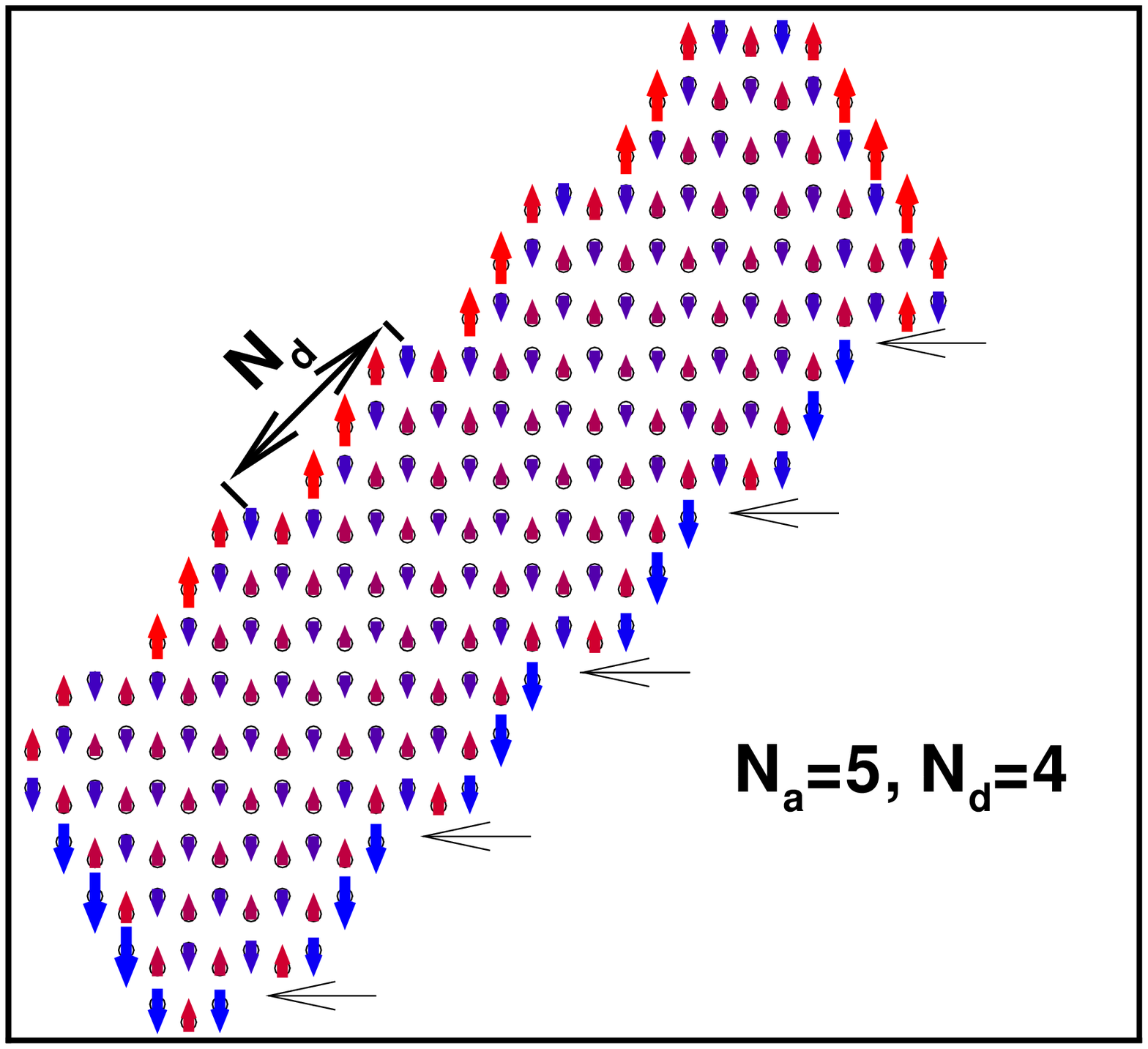}}
\subfigure[]{\epsfxsize=6.0truecm \epsfbox{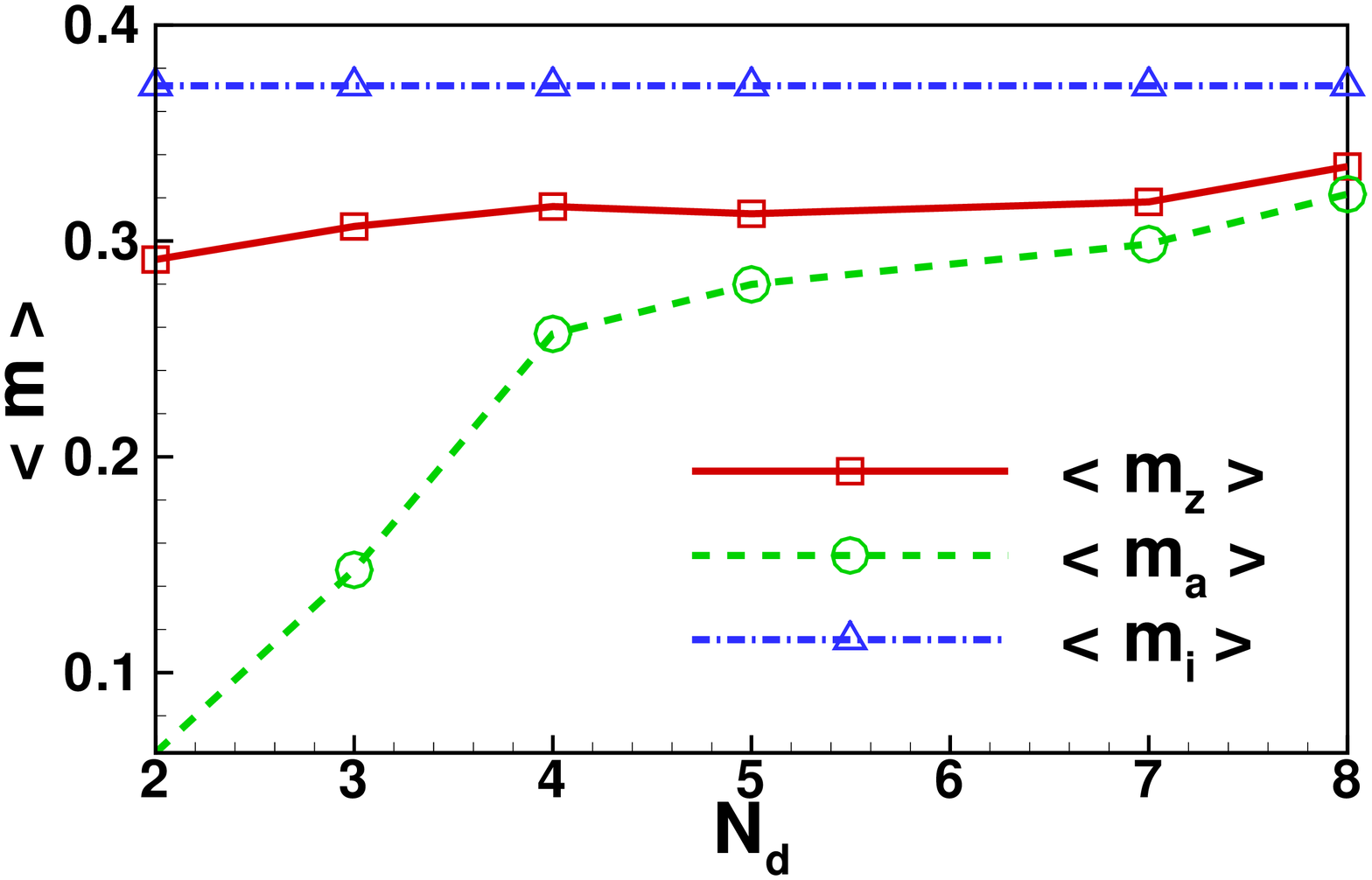}}
\caption{(Color online) Magnetic structure of graphene nanoribbons.
Nanoribbon in (a) is enclosed by the zigzag edges. The long
zigzag edges of the nanoribbons (b) and (c) is broken by intermittent
armchair ``defects''. As we decrease $N_d$, the interval
between the armchair ``defects'' (i.e., increase $N_a$, the total number of armchair
defects along the long side), the magnitude
of the average localized magnetic moment along the long side (represented by $\langle m_a\rangle$)
decreases. But $\langle m_z\rangle$, average localized moment along
the short side remains unaffected. We denote average moment per edge site of the ideal zigzag
structure (illustrated in (a)) by $\langle m_{i}\rangle$ in (d). Note the sharp drop of
$\langle m_a\rangle$ below $N_d=4$, which is the limit of three zigzag repeat units.}
\label{rand1}
\end{figure*}
%=========================================================================
\subsection{Nanoribbons}

The nanodots discussed in the last section had ``undisturbed'' zigzag
edges. It is interesting to investigate if ``defects'' present in a
zigzag edge can destroy the magnetism. We investigate this issue by
considering nanoribbons which have a large aspect ratio. We start with
nanoribbons enclosed solely by the zigzag edges, and introduce ``armchair
defects'' along the long edges of the nanoribbon (see
\fig{rand1}). The number of armchair defects introduced is denoted by
$N_a$ while the spacing between these defects is denoted by $N_d$.
Therefore if the value of $N_d$ is smaller, the number of armchairs
$N_a$ is higher and vice versa. Here we define $\langle m_z \rangle$
as the average magnitude of the magnetic moment along the short edges,
$\mean{m_a}$ as the average magnetic moment along the long edges
possibly containing armchair defects. \Fig{rand1}(d) shows a plots of
these quantities.  We observe from \fig{rand1}(d) that the moment
along the short sides (which are defect free) is essentially
unaffected by the number of defects on the long side. However
$\mean{m_a}$ falls gradually with the decrease of $N_d$ (increase of the number
of the armchair defects) until $N_d$ reaches about $4$, where there is a
dramatic fall in $\mean{m_a}$. Interestingly, this value of $N_d$
corresponds the situation where the contiguous zigzag segments in the
``defected zigzag edge''  have about three to four repeat units.
We therefore find a result that is fully consistent with that found in
the case of regular polygon nanodots; as long as there are three to
four repeat units of zigzag edge present, the ferromagnetism on
the edge is robust.

%fig======================================================================
\begin{figure*}
\subfigure[]{\epsfxsize=8.0truecm \epsfbox{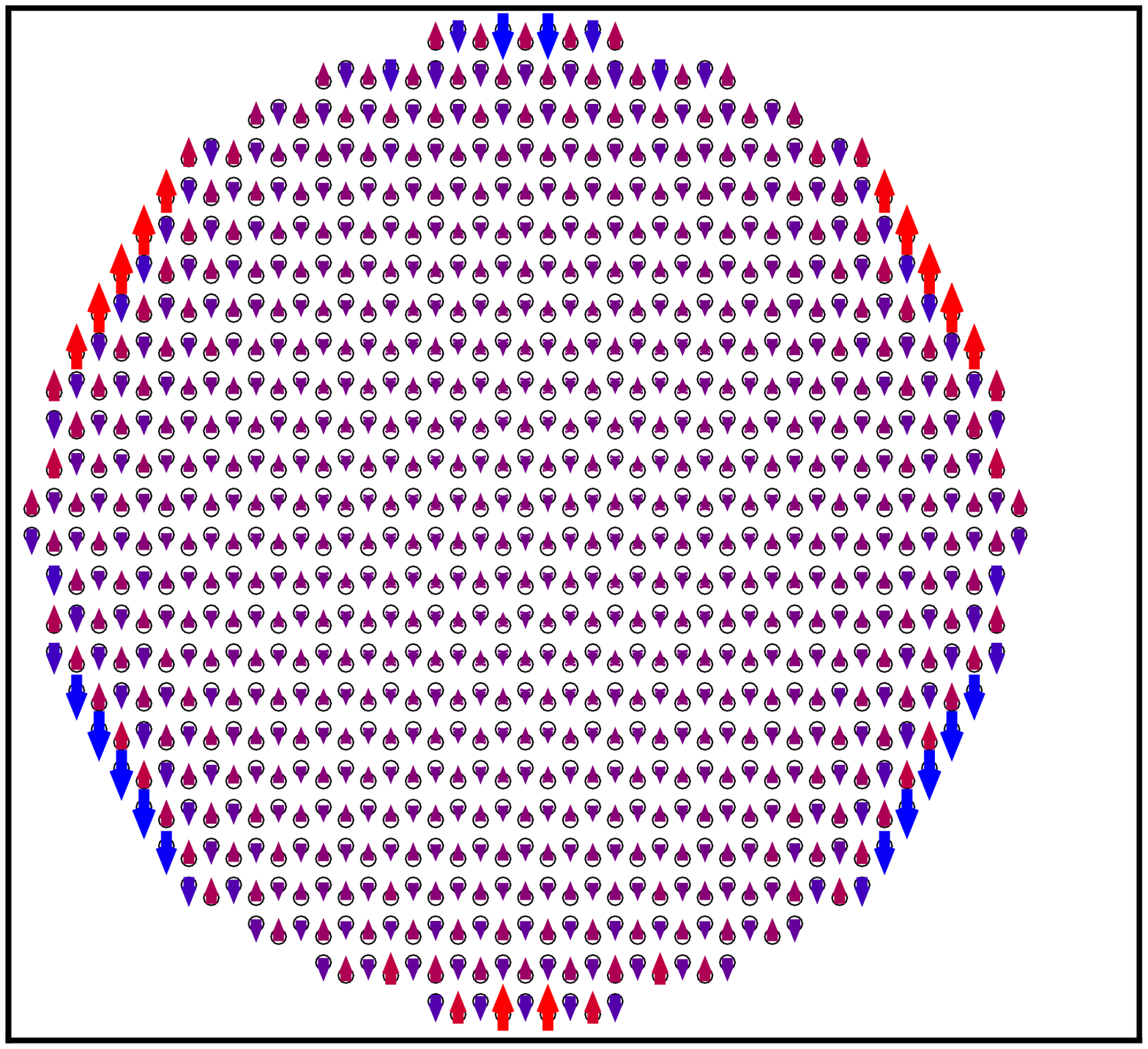}}
\subfigure[]{\epsfxsize=8.0truecm \epsfbox{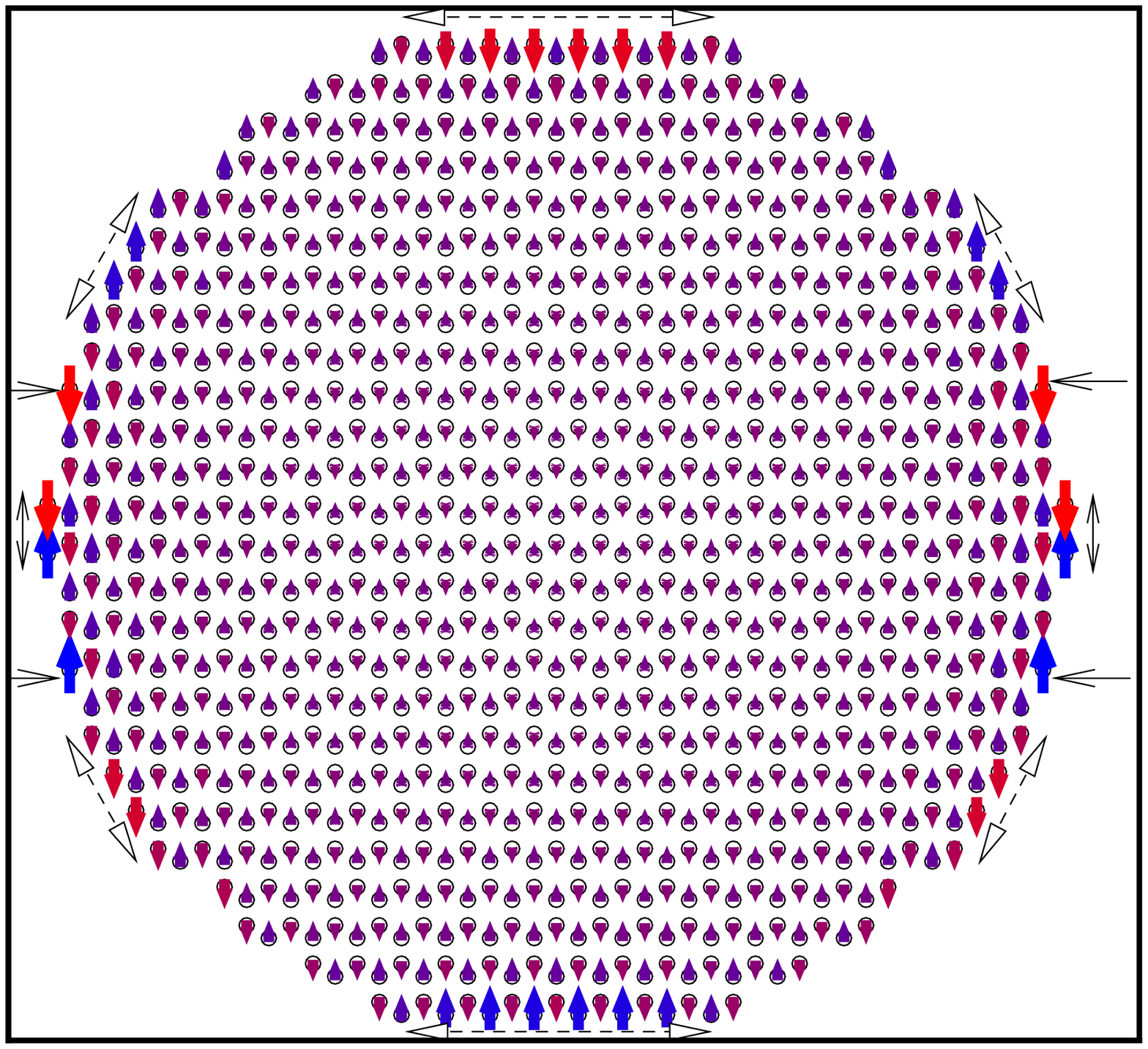}}
\subfigure[]{\epsfxsize=8.0truecm \epsfbox{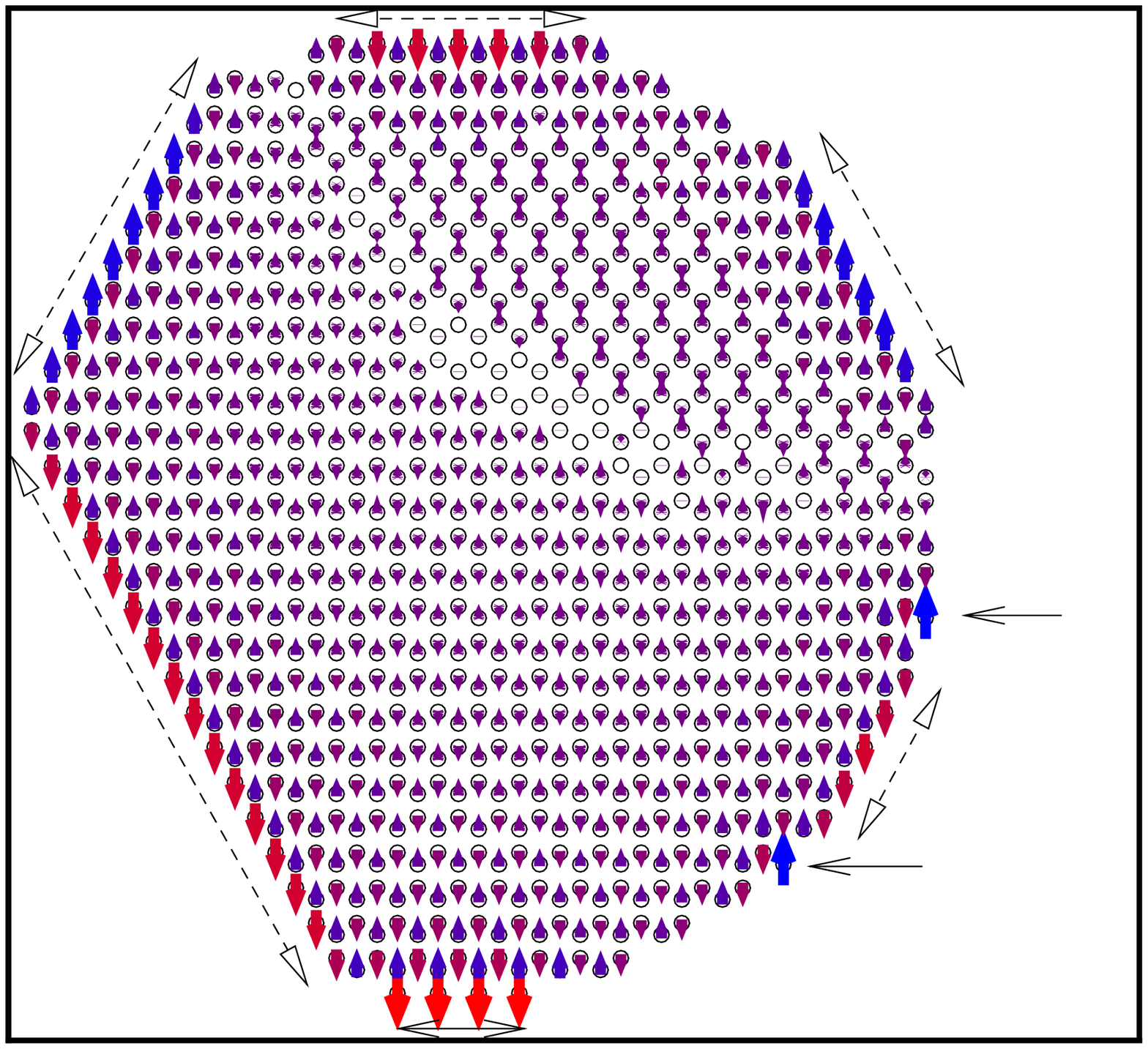}}
\subfigure[]{\epsfxsize=8.0truecm \epsfbox{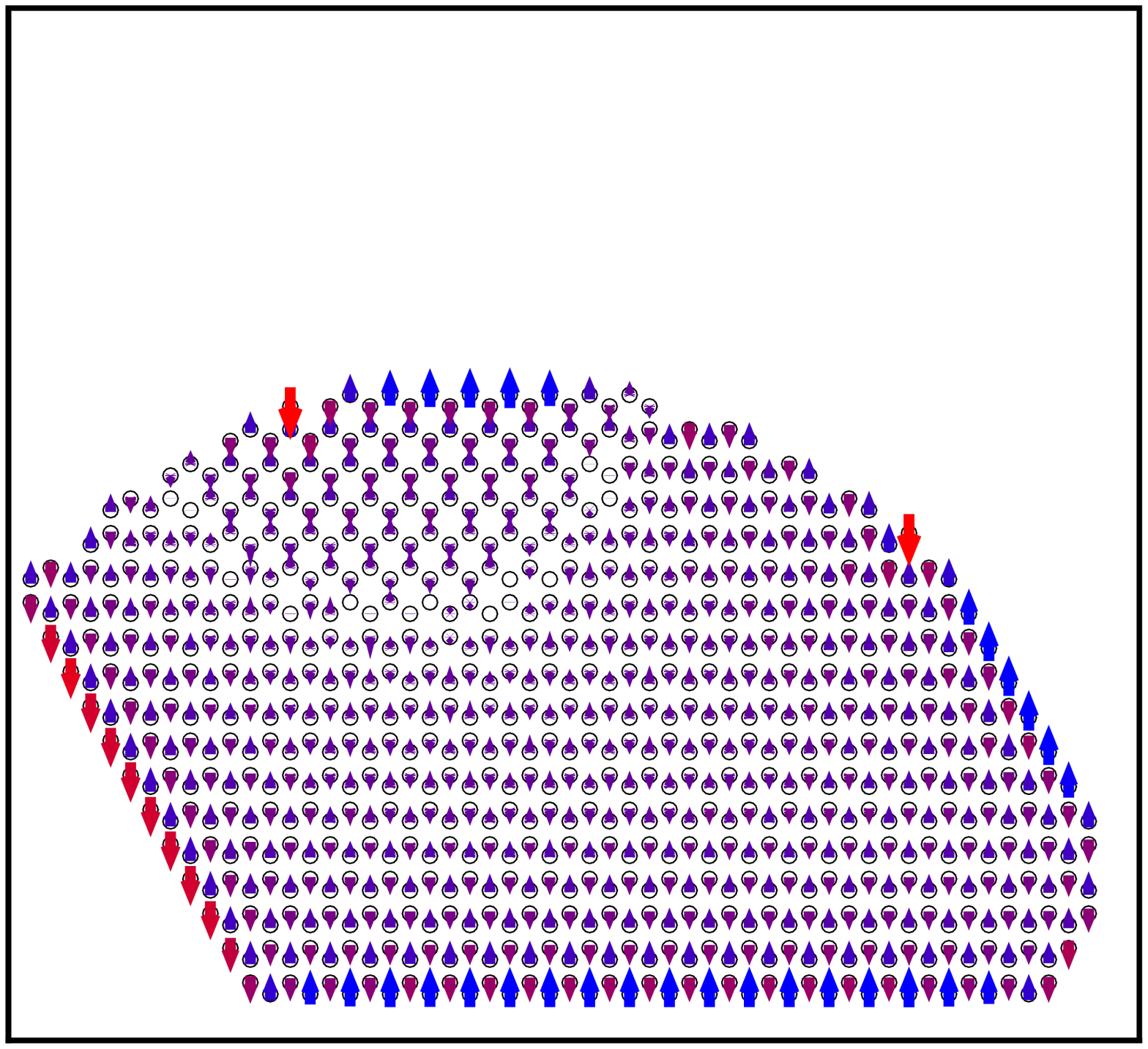}}
\caption{    (Color online) Magnetic structure of irregular graphene nanodots. (a),(b)  ``Circular cut-out'' nanodots. There are two different type of sites where the magnetic moment is high. First,
the usual zigzag edges, lying along the horizontal side and marked by dashed arrow. Second, a  set
of sites, marked by solid arrows, have only one nearest neighbor, as opposed to two in case of zigzags
and armchair edges. The second set of sites have even higher magnetic moment than compared to the zigzag
sites. (c),(d) - nanodots with both regular edges and a ``circular boundary''.
}
\label{rand2}
\end{figure*}
%===================================================================================

\subsection{``Irregular'' Nanostructures}

We shall now investigate several ``irregular'' nanodots created by
``cutting out'' patches out of a graphene sheet. Such a
random cutout can produce edge atoms which have two nearest
neighbours, or a single nearest neighbour. In the latter case the
structure thus obtained may not be an equilibrium structure; we ignore
this complication for the present study and focus on the nature of edge
magnetism that arises in irregular structures. A study of \fig{rand2}
shows that the edge magnetism is very robust even in irregular
structures; we find again that as long as there are three to four
repeat units of a zigzag edge present, the edge magnetism is
significant. We find an additional interesting feature. The sites
with a single nearest neighbour (such as those that are indicated by
a solid arrow in \Fig{rand2}) have very large magnetic moments, larger
than even the zigzag edges. As noted above, these are the ``high energy
edges'' and are not likely to be supported by thermodynamics.

\section{Conclusion}
\label{conclusion}
As noted in the introduction there have been a large number of
theoretical efforts investigating edge state
magnetism\cite{okada2001,jiang2007,rossier2007,castrocondmat2008,katsunori},
in {\it regular} nanostructures terminated by zigzag edges.  However,
we are not aware of any direct experimental observation of this
phenomena for graphene nanoribbons or nanodots. There are a few
experimental reports of magnetism in activated carbon
fibers~\cite{shibayama2000,toshiaki2005}, composed of a disordered
network of nanographite, which remains to be the best proof of the
theoretical prediction so far.  In this paper we have enquired if
``irregularity'' of the experimental graphene nanostructures could be
behind the lack of direct experimental corroboration. We have
investigated how the edge state magnetism of the zigzag edges is affected
by the presence of other edges, ``defects'' and random terminations.
We find that the edge state magnetism is very robust to these
``mutilations'' of the nanostructures -- \addn{for a value of $U=2 t$, as long as there are three to four
repeat units of a zigzag edge, the edge state magnetism is
preserved. For smaller values of on site Coulomb repulsion, more number of repeat units of a
zigzag edge is required for sustaining edge state magnetism (e.g., for $U=1.2 t$, the critical number of repeat units
is five to six).} In addition, we note that certain ``high energy'' edges
(ones where the edge atoms have only one nearest neighbour) can have
very large moments compared to even the zigzag edges. 
Thus our study
 demonstrates that the shape irregularity is unlikely to
destroy the edge state magnetism. However, it is clear that atomic resolution magnetic force
microscopy may have to be employed for observation of edge state magnetism in short zigzag
segment of graphene.\cite{kaiser2007} On the theoretical front, the present approach is 
based on the mean field theory which, of course, neglects quantum fluctuation effects. These may
be expected to be important in an ``effectively 1D'' system like the zigzag edge of graphene and
have to be investigated further.

Note added: After completion of the manuscript, we became aware of a similar work by H. Kumazaki
\etal\cite{kumazaki2008} We thank Kumazaki \etal for bringing this to our notice.

\section{Acknowledgment}
The authors thank G.~Baskaran, R.~Shankar and U.~Waghmare for
discussions. The authors are particularly grateful to A.~Ghosh for
discussions, suggestions and comments.  VBS thanks DST for generous
support for this work through a Ramanujan grant.

%-----------------------------------------------------------------------------
%BIBLIOGRAPHY ~/cite{key}_____________________________________________________
%\bibliographystyle{prsty}
\bibliography{ref.bib}
%\begin{thebibliography}{99}
%\end{thebibliography}
\end{document}